# Fully automated spectroscopic ellipsometry analyses: Application to MoO$_x$ thin films

Kohei Oiwake,[a] Yukinori Nishigaki,[a] Shohei Fujimoto, Sara Maeda, and Hiroyuki Fujiwara[b]

Department of Electrical, Electronic and Computer Engineering, Gifu University, 1-1 Yanagido, Gifu 501-1193, Japan.

Abstract

In spectroscopic ellipsometry, the optical properties of materials are obtained indirectly by generally assuming dielectric function and optical models. This ellipsometry analysis, which typically requires numerous model parameters, has essentially been performed by a try-and-error approach, making this method as a rather time-consuming characterization technique. Here, we propose a fully automated spectroscopic ellipsometry analysis method, which can be applied to obtain dielectric functions of light absorbing materials in a full measured energy range without any prior knowledge of model parameters. The developed method consists of a multiple-step grid search and the following non-linear regression analysis. Specifically, in our approach, the analyzed spectral region is gradually expanded toward higher energy while incorporating an additional optical transition peak whenever the root-mean-square error of the fitting analysis exceeds a critical value. In particular, we have established a unique algorithm that could be employed for the ellipsometry analyses of different types of optical materials. The proposed scheme has been applied successfully for the analyses of MoO$_x$ transparent oxides and the complex dielectric function of a MoO$_x$ layer that exhibits dual optical transitions due to band-edge and deep-level absorptions has been determined. The developed method can drastically reduce a time necessary for an ellipsometry analysis, eliminating a serious drawback of a traditional spectroscopic ellipsometry analysis method.

[a] Contributed equally to this study.
[b] fujiwara@gifu-u.ac.jp

## I. INTRODUCTION

Spectroscopic ellipsometry (SE) is essentially an indirect characterization method, in which the optical constants (i.e., refractive index $n$ and extinction coefficient $k$) are determined from fitting of calculated SE spectra to experimentally measured spectra.[1,2] This fitting analysis, which typically includes modeling of sample dielectric functions ($\varepsilon = \varepsilon_1 - i\varepsilon_2$), has been performed manually based on human intuition and is extremely time consuming. Specifically, SE fitting analyses generally require numerous variables for expressing sample optical properties and structures[1-3] and enormous efforts are necessary to obtain full optical function spectra in a measured region, even though SE measurements themselves take only a few seconds.[1] In essence, indirect SE tends to be a hard-to-analyze method, which requires certain knowledge and experience for modeling optical characteristics and structures. As a result, in many cases of material optical characterization, the SE technique has been avoided and a simple transmittance/reflectance (T/R) technique has more popularly been adopted,[4-6] although accuracy of SE is far better than the simple T/R method,[7,8] which is based on a highly hypothetical and idealized analysis. [4,5]

The complex SE analyses particularly for thin film samples can be avoided if we can perform mathematical inversion (or point-by-point fitting), where $(n, k)$ spectra are obtained directly from ellipsometry ($\psi, \Delta$) spectra using structural parameters deduced from the analysis of a thin-film optical interference.[1,2] Just recently, machine learning approach has been proposed to automate this inversion problem.[9] Nevertheless, sample imperfection and non-ideal surface/interface structures often prevent the direct assessment of material optical constants by mathematical inversion and many SE analyses have been performed by expressing sample dielectric functions based on optical-transition peak modeling.[1,2,10,11] To date, a fully automated SE analysis method incorporating dielectric function modeling has not been developed. Such an automated SE analysis is expected to open a new opportunity for SE as a fast and reliable characterization technique for the optical-constant determination of a unique variety of materials including semiconductors, insulators, metals, transparent conductive oxide (TCO) and organic/bio materials.

In this study, we have developed the first fully automated SE analysis method with dielectric function modeling, in which the ellipsometry analyses are performed by gradually expanding an analyzed energy region toward higher energy, while incorporating additional transition peaks to express complex light absorption phenomena in materials. The SE automation has been realized by further incorporating a multi-step grid search for obtaining initial guess values for effective SE fitting. Our approach has been applied successfully to analyze $MoO_x$ transparent oxides, demonstrating completely automated SE characterization without any prior knowledge of structural and material parameters. The developed scheme can potentially be adopted to perform SE analyses for all possible materials, for which light reflection can be measured.

## II. AUTOMATED SE ANALYSIS METHOD

### A. Principle of the proposed method

Figure 1 explains the procedure of a newly developed SE analysis. In the proposed method, the SE spectra ($\tan\psi$, $\cos\Delta$) are analyzed consecutively from a low energy region. For this approach, the material dielectric function is expressed by combining multiple transition peaks described by the Tauc-Lorentz (TL) model.[12] In the initial analysis performed in a low energy region, only one TL peak is assumed and the highest analyzed energy ($E_{HA}$) is expanded gradually with a step of $\Delta E$ while implementing a non-linear SE fitting analysis repeatedly. However, the optical function of many materials cannot be described by a single TL peak[3] and the root-mean-square error in the fitting ($\sigma$), which shows a difference between experimental and calculated ellipsometry spectra, increases gradually with increasing $E_{HA}$ (see Fig. 1). When $\sigma$ reaches a critical value ($\sigma_{crit}$), a new TL peak is added and the fitting quality improves by the incorporation of additional fitting parameters. The above process is repeated while adding a new transition peak whenever $\sigma$ reaches $\sigma_{crit}$ until $E_{HA}$ becomes equivalent to the maximum measured energy (i.e., $E_{HA} = E_{max}$). This is the principle of the proposed method.

In earlier manual SE analyses, the variation of an analyzed energy range (or a wavelength range) was also adopted to determine a fine-structured dielectric function in an infrared region[13] or light-penetration-depth-dependent thin film structures.[14,15] In particular, for the infrared SE analysis, the analyzed region was expanded toward lower energy (i.e., longer wavelength) to resolve sharp absorption features observed in the infrared region.[13]

In our method, established to analyze the SE spectra in the visible/ultraviolet region, the $\varepsilon_2$ spectrum of samples is modeled as a sum of several TL peaks[3,12]:

$$\varepsilon_2(E) = \sum_{j=1}^{N} \frac{A_j C_j E_{0,j} (E - E_{g,j})^2}{[(E^2 - E_{0,j}^2)^2 + C_j^2 E^2]E} . \qquad (1)$$

A single TL peak is expressed by four parameters: namely, the amplitude parameter $A_j$, broadening parameter $C_j$, peak transition energy $E_{0,j}$, and optical gap $E_{g,j}$.[12] From the parameters of Eq. (1), $\varepsilon_1(E)$ can be calculated based on Kramers-Kronig integration of each TL peak:

$$\varepsilon_1(E) = \varepsilon_1(\infty) + \sum_{j=1}^{N} \left( \frac{2}{\pi} P \int_{E_g}^{\infty} \frac{E' \varepsilon_{2,j}(E')}{E'^2 - E^2} dE' \right), \qquad (2)$$

where $\varepsilon_1(\infty)$ represents a constant contribution to $\varepsilon_1(E)$ at high energies. The second term on the right side shows Kramers-Kronig integration, which can be performed using the exact equation.[12]

We have previously established that the TL model can be applied to express the dielectric functions of a wide variety of materials, including semiconductors, TCOs, metals, and organic materials[3]; thus, the TL model is sufficient for expressing various material optical properties, although the Drude model is necessary when free carrier absorption is present.[2,16] When $\varepsilon_2$ spectra are parameterized using several TL peaks, however, the material optical functions are defined by multiple $E_g$ values [i.e., $E_{g,j}$ in Eq. (1)]. The lowest $E_{g,1}$ value (i.e, $j = 1$) may correspond to the material $E_g$, but the other $E_{g,j}$ values are employed solely for the parameterization and do not have distinct physical meaning.

For the calculation of $\sigma$, the following general equation has been adopted[1]:

$$\sigma = \frac{1}{\sqrt{2M - P}} \left\{ \sum_{i=x}^{M} \left( [\tan \psi_{ex}(E_i) - \tan \psi_{cal}(E_i)]^2 + [\cos \Delta_{ex}(E_i) - \cos \Delta_{cal}(E_i)]^2 \right) \right\}^{1/2} , \qquad (3)$$

where $M$ and $P$ show the numbers of measurement points in ellipsometry spectra and analytical parameters, respectively. The subscripts of "ex" and "cal" represent experimental and calculated values at the energy of $E_i$. In the developed method, the SE fitting is performed in the region of $E_i \leq E_{HA}$. However, $\sigma$ is calculated using only a high energy part of the fitted spectra. Specifically, $\sigma$ is determined in a selected energy region of $E_{HA} - \Delta E_{HA} \leq E_i \leq E_{HA}$, where $\Delta E_{HA}$ indicates an energy range used for the $\sigma$ calculation ($\Delta E_{HA} = 0.5$ eV in this study). In our SE analysis, therefore, the total number of data points used for the $\sigma$ calculation is always the same. This measure is vital as $\sigma$ is a critical value used to judge the incorporation of an additional transition peak. In contrast, when $\sigma$ is determined from a whole fitting range, the increase in $\sigma$ tends to be suppressed, particularly when $(\tan\psi, \cos\Delta)$ fitting is better in the low energy region. Moreover, to maintain high $\sigma$ sensitivity in the fitting analysis, we generate experimental $\tan\psi_{ex}(E)$ and $\cos\Delta_{ex}(E)$ with an equal energy step of 0.01 eV by interpolating experimental data and these modified spectra were employed in the analyses, while the raw ellipsometry spectra have an equal spacing for wavelength. On the other hand, our SE simulations and the following analyses show that the SE fitting with $(\tan\psi, \cos\Delta)$ provides a higher accuracy, compared with $(\psi, \Delta)$ fitting, whereas the SE analyses using Stokes parameters (i.e., $S_1 = -\cos 2\psi$, $S_2 = \sin 2\psi \cos\Delta$, and $S_3 = -\sin 2\psi \sin\Delta$) yield similar accuracies to $(\tan\psi, \cos\Delta)$.

As the optical model for a thin-film layer formed on a substrate, we employed a general model of (surface roughness layer)/(bulk layer)/substrate with $d_b$ and $d_s$ being the bulk layer and surface roughness layer thicknesses, respectively.[1] In this case, the structure of the thin film is expressed simply by two values of $(d_b, d_s)$. The optical properties of the surface roughness layer are calculated from those of the bulk layer by applying effective medium approximation assuming a void volume fraction of 50 vol.% within the roughness layer.[1,2,17]

## B. Procedure of the proposed method

Figure 2 shows the flow chart of the developed automated SE analysis. In this approach, we first set $\sigma_{crit}$ and $E_{HA}$. In this study, the initial settings are $\sigma_{crit} = 1 \times 10^{-3}$ and $E_{HA} = 1.5$ eV. The lowest measured energy is $E_{min} = 0.75$ eV and thus the first SE fitting is performed at $0.75 \le E_i \le 1.5$ eV. Unfortunately, the SE fitting based on a non-linear regression analysis depends critically on initial guess parameters and, to obtain appropriate initial parameters, we implement a grid search, where initial values are changed sequentially with proper steps to find starting values that provide the lowest $\sigma$ within a searched grid. In particular, we have developed a multiple grid search scheme to reduce the computational cost, as described in next section.

In the first SE fitting performed using only one TL peak [$N = 1$ in Eq. (1)] with $E_{HA} = 1.5$ eV, the number of fitting parameters is seven ($n_{fp} = 7$): i.e., five parameters for the TL peak [$A_1$, $C_1$, $E_{0,1}$, $E_{g,1}$, and $\varepsilon_1(\infty)$] and two structural parameters ($d_b$ and $d_s$). Accordingly, the grid search is carried out in this seven-dimensional space. After the initial guess values are obtained, the non-linear regression analysis is performed using all the fitting parameters (i.e., all parameter fitting) with $E_{HA} = 1.5$ eV. This is followed by the expansion of the analyzed spectral region, expressed as $E_{HA} = E_{HA,old} + \Delta E$. Here, $E_{HA}$ is the extended energy with $\Delta E$ being the expanded energy step ($\Delta E = 0.1$ eV in this study). The SE fitting is performed again for the expanded spectral region and $\sigma$ for this fitting is evaluated. If $\sigma > \sigma_{crit}$, an additional TL peak is added and the grid search is performed again. However, when $N > 1$, the grid search is carried out for only additional parameters introduced by the incorporation of the new TL peak (i.e., $A_j$, $C_j$, $E_{0,j}$, and $E_{g,j}$), followed by all parameter fitting with $n_{fp} = 4N + 3$. Here, the three peak independent parameters are $\varepsilon_1(\infty)$, $d_b$, and $d_s$. The analyzed range is expanded further and the restricted SE fitting is performed for a higher $E_{HA}$. In this restricted SE fitting implemented for $N > 1$, only the TL peak having the highest transition peak energy (i.e., $E_0$) is adjusted, together with the structural parameters (i.e., $n_{fp} = 7$ in the restricted fitting). The expansion of the analyzed region and the following SE fitting are implemented continuously until $E_{HA}$ reaches $E_{max}$, while introducing a new TL peak whenever $\sigma$ exceeds $\sigma_{crit}$. The final SE result is obtained after performing all parameter fitting in the whole energy region.

In the developed method shown in Fig. 2, the restricted SE fitting with $n_{fp} = 7$ is vital to suppress the analysis error. It should be emphasized that SE fitting analyses incorporating the TL free parameters are essentially unstable and tend to generate fitting errors. Specifically, for the TL model, the following restrictions exist: i) $C < 2E_0$, ii) $E_g < E_0$, and iii) $A > 0$. Thus, the SE fitting results of $C > 2E_0$, $E_g > E_0$, and $A < 0$ define the failure of SE fitting analyses. When the result of $C > 2E_0$ was obtained, we fixed $C$ and $E_0$ values to those obtained in the earlier analysis with a smaller fitting region (i.e., $E_{HA,old}$ in Fig. 2). Similarly, in the case of $E_g > E_0$, the fixed $E_g$ and $E_0$ obtained in the $E_{HA,old}$ analysis can be adopted. Nevertheless, in cases summarized below, $\sigma_{crit}$ is increased by $\Delta\sigma$ ($\Delta\sigma = 1 \times 10^{-3}$ in this study):

i)       when $A < 0$,

ii)      when the analysis errors of $C > 2E_0$ and $E_g > E_0$ are observed simultaneously,

iii)     when the analysis error occurs even when an additional TL peak is added, or

iv)     when $\sigma$ does not decrease even when an additional TL peak is added.

When $\sigma_{crit}$ is increased, the whole routine is started from the beginning. Accordingly, $\sigma_{crit}$ should be set to a sufficiently low value initially ($\sigma_{crit} = 1 \times 10^{-3}$ in this study) and, whenever the fitting leads to an error, $\sigma_{crit}$ is increased until $E_{HA}$ reaches $E_{max}$.

## C. Grid search

The proposed method relies heavy on the accuracy of the grid search as there are many variables in SE analyses and inappropriate choice of the initial fitting values leads to serious fitting errors. The success of the grid search depends critically on a mesh size or a step size in the variation of initial guess values. Nevertheless, when the mesh size is too small (i.e., when the guess values are changed with small variation steps), the number of the total combination of fitting variables becomes too large, resulting in an enormous calculation time. For example, when the mesh size is $k = 10$ and each fitting parameter for $N = 1$ is divided into 10 initial guesses in a certain range, all the possible combinations of the TL peak and structural parameters reach $10^7$ patterns ($n_{\mathrm{fp}} = 7$). Although the restrictions of $C < 2E_0$ and $E_{\mathrm{g}} < E_0$ reduce the actual combinations, such analyses still take > 10 hr if a conventional computer is employed. Thus, the reduction of the mesh grid size is vital to implement the SE analysis within a manageable time scale.

To reduce the computational cost for the grid search, we have developed a multi-step grid search (zooming grid search). In this method, the first grid search is carried out using a coarse mesh to find the first set of the guess values and the second grid search is further performed by zooming the grid around the guess values obtained in the first grid search. The third grid search (final grid search) is also performed near the data set determined in the second grid search. Although a single-step grid search has been employed widely in ellipsometry data analyses,[1] the developed multi-step method is computationally quite effective; in particular, the appropriate initial values can be determined even in the seven-dimensional parameter space in a shorter time (~2 hr). In this study, the first grid search is carried out with a mesh size of $k = 5$, whereas the second and third grid searches are implemented with a denser mesh of $k = 7$.

In our approach, when the SE analysis is performed using only one TL peak ($N = 1$), the grid search is carried out for all the seven parameters [i.e., $A_1$, $C_1$, $E_{0,1}$, $E_{\mathrm{g},1}$, $\varepsilon_1(\infty)$, $d_{\mathrm{b}}$, and $d_{\mathrm{s}}$]. However, when a new TL peak is added to improve the fitting, the grid search is implemented only for the four parameters of the added TL peak (i.e., $A_j$, $C_j$, $E_{0,j}$, and $E_{\mathrm{g},j}$). This procedure reduces the computational time greatly and the grid search can be performed within ~5 min.

For $N = 1$, the actual ranges used for the first grid search are i) $A_1 = 0.5{\sim}200$ eV, ii) $C_1 = 0.25{\sim}3$ eV, iii) $E_{0,1} = 0.1{\sim}2.5$ eV, iv) $E_{\mathrm{g},1} = 0.1{\sim}2.0$ eV, v) $\varepsilon_1(\infty) = 1{\sim}6$, vi) $d_{\mathrm{b}} = 10{\sim}200$ nm, and vii) $d_{\mathrm{s}} = 10{\sim}50$ nm. These ranges, however, need to be modified slightly depending on the material system. The variable ranges of the second grid search are expressed by $0.5G_{\mathrm{1st}} \leq F \leq 1.5G_{\mathrm{1st}}$, where $F$ and $G_{\mathrm{1st}}$ show the fitting parameter and a guess value determined in the first grid search. For example, for $E_{0,1}$, the first grid search is performed using a five data set ($k = 5$) of $E_{0,1} = 0.1$, 0.7, 1.3, 1.9, and 2.5 eV. If the lowest $\sigma$ is obtained at the grid of $E_{0,1} = 1.3$ eV, the second grid search is carried out in a range of $0.5 \times 1.3 \leq E_{0,1} \leq 1.5 \times 1.3$ (i.e., $F = E_{0,1}$ and $G_{1st} = 1.3$ eV) with $k = 7$. The third grid search ($k = 7$) is further performed in a similar manner in the range of $0.5G_{2nd} \leq F \leq 1.5G_{2nd}$, where $G_{2nd}$ represents the guess value deduced in the second grid search.

In our zooming grid search, a rather wide range covering from 50% to 150% of the grid search result is used. This is necessary as the variable ranges of the first grid search are not wide enough and the best $\sigma$ could be obtained at the boundary of the grid. For example, if $A_1 = 200$ eV is obtained as the best initial guess in the first grid search, $A_1 = 1.5 \times 200 = 300$ eV is possible in the second grid search, which could further be extended to $A_1 = 450$ eV in the third grid search. Accordingly, even though the first grid search is performed in relatively narrow ranges, larger changes are possible in our zooming grid search.

For $N \geq 2$, the optical transition energies (i.e., $E_0$ and $E_g$) of an added peak become larger, as $E_{HA}$ increases. Thus, the initial ranges of $E_{0,1}$ (0.1~2.5 eV) and $E_{g,1}$ (0.1~2.0 eV) assumed for $N = 1$ could be too small and need to be increased. In our analysis for $N \geq 2$, therefore, a slightly higher range of $E_{0,N} = 1.2$~3.7 eV was adopted. The $E_{g,N}$ was also controlled as $0.5E_{HA} \leq E_{g,N} \leq 1.5\ E_{HA}$.

## III. EXPERIMENT

### A. MoO$_x$ deposition

MoO$_3$ is a wide gap $n$-type oxide with $E_g = 2.8$ eV (Ref. 18) and MoO$_x$ ($x \leq 3$) layers have been applied widely to solar cell devices as a high work-function material.[19-21] The SE analyses of MoO$_x$ have already been reported.[22-24] In this study, MoO$_x$ layers with a thickness of 80 nm were fabricated on glass substrates (Eagle XG) by rf sputtering with a rf power of 40 W at room temperature using a Mo target. When MoO$_x$ is fabricated at room temperature, the structure becomes amorphous,[23,25] which can be crystallized at ~350 °C by thermal annealing.[25]

We have fabricated two different MoO$_x$ samples (i.e., O-rich and O-poor samples) by varying the ratio of Ar and O$_2$ flow rates while maintaining the same Ar flow rate of [Ar] = 40 SCCM and total pressure of 1.0 Pa. For the O-rich sample (MoO$_x$-A), a ratio of [Ar]/[O$_2$]=4 is used and the film composition analysis of the bare sample of MoO$_x$-A by x-ray photoelectron spectroscopy results in MoO$_{2.71}$. For the fabrication of the O-poor sample (MoO$_x$-B), a much higher ratio of [Ar]/[O$_2$] = 27 is adopted. When [O$_2$] is low, the oxygen deficiency creates mid-gap states[18,25] and the optical function of MoO$_x$ varies significantly.[23,25] In this study, the two different dielectric functions of MoO$_x$-A and MoO$_x$-B samples were analyzed.

## B. SE measurements and analyses

The ellipsometry spectra of MoO$_x$-A and MoO$_x$-B samples were measured by a rotating-compensator instrument (J. A. Woollam, M-2000DI). In these SE measurements, an incident angle was fixed at 55°. To suppress the light reflection from the glass rear surface, the Scotch tape was pasted onto the backside of the glass substrate.[26] For the SE analyses of MoO$_x$/glass structures, we employed the optical constants of a Eagle XG glass substrate, which is described by a formula of

$$n(\lambda) = A + \frac{B}{\lambda^2} + \frac{C}{\lambda^4} - D\lambda^2, \tag{4}$$

with $A = 1.503$, $B = 2876$ nm$^2$, $C = 8.852 \times 10^7$ nm$^4$, $D = 5.196 \times 10^{-9}$ nm$^{-2}$. The dielectric functions of the MoO$_x$ layers were determined by the automated SE analyses described in Section II.

## IV. RESULTS

### A. SE analysis of MoO$_x$-A

Figure 3 shows the result obtained for the MoO$_x$-A sample based on the developed automated SE analysis: (a) $\sigma$, (b) tan$\psi$, (c) cos$\Delta$, (d) $\varepsilon_2$, and (e) $d_b$ and $d_s$ values. In Fig. 3(a), the variation of $\sigma$ with $E_{HA}$ is shown and each $\sigma$ is calculated from Eq. (3) with $\Delta E_{HA} = 0.5$ eV. In the initial analysis with $E_{HA} = 1.5$ eV, the tan$\psi$ and cos$\Delta$ spectra in the range of 0.75~1.5 eV were fitted using only one TL peak ($N = 1$) and the orange lines in Figs. 3 (b) and (c) show the result of the zooming grid search and the following all parameter fitting ($n_{tfp} = 7$). The resulting $d_b$ and $d_s$ are also shown in Fig. 3(e). When $E_{HA} = 1.5$ eV, $\sigma$ is sufficiently low ($\sigma = 3.69 \times 10^{-4}$).

In the developed method, the analyzed energy range (i.e., $E_{HA}$) is gradually expanded. However, the SE fitting with one TL peak becomes increasingly difficult and $\sigma$ gradually increases as $E_{HA}$ shifts toward higher energy. In the analysis of MoO$_x$-A, the optimum $\sigma_{crit}$ is found to be $3 \times 10^{-3}$ and the second TL peak is added at $E_{HA} = 4.3$ eV when $\sigma > \sigma_{crit}$. The SE fitting with two TL peaks (i.e., $N = 2$) improves the fitting quality notably and $\sigma$ decreases from $3.35 \times 10^{-3}$ to $6.29 \times 10^{-4}$ at $E_{HA} = 4.3$ eV. The black lines in Figs. 3(b)-(d) show the spectra obtained after the zooming grid search and all parameter fitting for $N = 2$. As $E_{HA}$ is extended further, $\sigma$ increases again even with $N = 2$ and the third TL peak is incorporated into the analysis at $E_{HA} = 4.9$ eV and the optimized spectra obtained with $N = 3$ are shown by blue lines in Figs. 3(b)-(d). The same procedure is repeated and the result for $N = 4$ is shown by red lines. The final result is obtained by performing all parameter fitting with $N = 4$ at $E_{HA} = E_{max} = 6$ eV, as indicated by green lines. The $d_b$ and $d_s$ values obtained in this last fitting analysis give final results [$d_b = 81.7$ nm and $d_s = 4.8$ nm in Fig. 3(e)]. In this manner, the dielectric function of the whole measured range and structural parameters of the sample can be determined.

In the analysis of MoO$_x$-A, $\sigma_{crit} = 3 \times 10^{-3}$ allows the successful SE analysis, as mentioned above. When $\sigma_{crit}$ is set to a lower value (i.e., $< 3 \times 10^{-3}$), additional TL peaks are incorporated at a lower energy. In this case, however, the incorporation of the additional TL peak does not improve the fitting and the larger number of free parameters leads to fitting errors categorized in Fig. 2. Consequently, the quality of the automated SE analysis can be judged from the absolute value of $\sigma_{crit}$ obtained from the analysis.

Figure 4 shows the dielectric functions of MoO$_x$-A decided from (a) the automated SE analysis of Fig. 3 and (b) a manual SE analysis performed separately (black lines). In the manual analysis, $\sigma$ for the whole spectra with $M$ measurement points (i.e., $\sigma_M$) is lowered by a traditional trial-and-error approach assuming two TL peaks. In Fig. 4, the contributions of each TL transition peak are also indicated by colors. In the automated-SE result, the dielectric function of MoO$_x$ is described by four peaks and the result shows clear shoulder peaks at $E = 3.87$ and $4.77$ eV. In contrast, the precise fitting is extremely difficult in the manual SE and the strong absorption in the ultraviolet region is expressed by a single TL peak. As a result, $\sigma_M$ of the manual SE analysis ($2.60 \times 10^{-3}$) becomes larger, compared with $\sigma_M = 9.34 \times 10^{-4}$ obtained in the automated SE analysis. Accordingly, the developed method allows a more accurate determination of material optical properties. However, if $\sigma_{crit}$ in the automated SE analysis is increased, the fitting quality degrades and the resulting dielectric function becomes more similar to that obtained from the manual SE analysis. Thus, $\sigma_{crit}$ needs to be increased gradually to maintain high accuracy.

The numerical TL parameters extracted from the automated and manual SE analyses of Figure 4 are summarized in Supplementary Tables I and II. It can be confirmed that the 90% confidence limits of the TL parameters obtained for the automated analysis are larger than those determined from the manual SE analysis, as the number of the TL peaks is larger in the automated analysis, even though $\sigma_M$ of the automated analysis is smaller than the manual analysis.

**B. SE analysis of MoO$_x$-B**

Figure 5 shows the result of the automated SE analysis performed for MoO$_x$-B, which exhibits a more complex dielectric function due to the mid-gap-state formation. In Fig. 5, (a) $\sigma$, (b) $\tan\psi$, (c) $\cos\Delta$, (d) $\varepsilon_2$, and (e) $d_b$ and $d_s$ are shown in a similar manner with Fig. 3. The initial SE analysis is performed in a range of 0.75~1.5 eV and the result of the zooming grid search and all parameter fitting with $E_{HA}$ = 1.5 eV is shown by orange lines. As confirmed from Fig. 5(d), the MoO$_x$-B sample shows notable light absorption with a peak transition energy of 1.2 eV because of the gap-state formation.[18,25] For the SE analysis of MoO$_x$-B, the optimum $\sigma_{crit}$ is 9 × 10$^{-3}$ and the second TL peak is introduced at $E$ = 2.9 eV. The SE fitting result at $E_{HA}$ = 2.9 eV with $N$ = 2 is indicated by black lines in Figs. 5(b)-(d). The incorporated second TL peak expresses the band-edge absorption of the MoO$_x$ and the $\varepsilon_2$ increases gradually as $E_{HA}$ increases. The third and fourth TL peaks are further incorporated at $E_{HA}$ = 5.8 and 6.0 eV (= $E_{max}$), respectively. The SE fitting results for $N$ = 3 and 4 are shown by blue and red lines, respectively, and the structural parameters are also determined ($d_b$ = 81.5 nm and $d_s$ = 4.2 nm). The above result shows clearly that the fully automated SE analysis is possible even when the dielectric function of the sample shows complex transitions.

Figure 6 shows the dielectric function of MoO$_x$-B obtained from the SE analysis of Fig. 5. From the contributions of the four TL peaks indicated in this figure, it can be seen that the optical transition by the mid-gap states is expressed by one TL peak, whereas the interband transition is modeled by three TL peaks. In the analysis of MoO$_x$-B, however, the two shoulder peaks observed in MoO$_x$-A (see Fig. 4) are not reproduced because $\sigma_{crit}$ of MoO$_x$-B (9 × 10$^{-3}$) is much higher than that of MoO$_x$-A (3 × 10$^{-3}$) and the detailed fitting is not possible. In other words, the imperfection of the gap-state modeling contributes to increase $\sigma_{crit}$ and, due to this high $\sigma_{crit}$, the fitting quality in the high energy region is sacrificed. This effect can be evidenced in the poor fitting observed at $E \geq$ 4.2 eV in Fig. 5(b).

It should be noted that, for the case of MoO$_x$-B, the manual SE analysis provides a better result. Figure 7 compares (a) the SE fitting results and (b) the $\varepsilon_2$ spectra of MoO$_x$-B obtained from the automated and manual analyses. The manual SE analysis was implemented assuming three TL peaks, while four TL peaks were assumed in the automated analysis (see Supplementary Tables III and IV for the exact TL parameter values). Even though the larger number of TL peaks is incorporated in the automated analysis, the fitting quality of the automated analysis is poor in the ultraviolet region, compared with the manual analysis [see Fig. 7(a)], and $\sigma_M$ of the automated SE ($3.04 \times 10^{-3}$) is larger than that of the manual analysis ($\sigma_M = 1.93 \times 10^{-3}$). As described above, in the proposed method, the precision of overall SE fitting is governed by $\sigma_{crit}$ and, once $\sigma_{crit}$ increases in a lower energy, the high fitting quality cannot be maintained in the following high energy analysis. In contrast, in the case of the manual analysis, the overall $\sigma$ is not determined by the fitting in a particular energy region and the fitting in the higher energy region can be carried out independently. However, we mention that manual analyses that use a large number of fitting parameters are very time consuming and difficult particularly when the dielectric function shows a complex structure. In Fig. 7(b), although the manual-SE result is more accurate, the automated-SE result essentially reproduces the manual-SE result and thus the fully automated SE analysis can still be performed for the $MoO_x$-B sample.

## V. DISCUSSION

In the application of ellipsometry to Si-based LSI technologies, material optical properties and detailed device structures are known quite well;[2,3,27] for such a research area, the impact of the developed method is less significant. In contrast, the automated ellipsometry analysis method proposed in this study becomes quite important when samples with unknown optical properties are analyzed, particularly by some who have less experience and knowledge for SE fitting analyses.

The dielectric functions of $MoO_x$ layers analyzed in this study show a broad feature due to the formation of an amorphous phase. In contrast, crystalline materials generally exhibit complex sharp features[3] and, at this stage, the applicability of the developed method for such crystal-phase materials is not clear. It should be emphasized that dielectric functions of almost all materials can be expressed assuming the TL transitions.[3] Thus, it is likely that our automated SE analysis can be applied to the SE analyses of other materials. It also needs to be mentioned that the positions and widths of transition peaks can be deduced directly from experimental $\langle \varepsilon_2 \rangle$ spectra (i.e., pseudo-dielectric function). The peak fitting analysis performed in advance for $\langle \varepsilon_2 \rangle$ can be effective in extracting guess parameters, which can then be adopted in the following automated SE analyses. Such a pre-fitting assessment could be essential in analyzing crystalline samples exhibiting complex absorption features.

We mention that, instead of the TL model used in this study, other transition models,

such as Lorentz, Gaussian, critical-point, and spline models,[2] can also be adopted in the automated SE analyses. For the Kramers-Kronig integration of $\varepsilon_2$ peak models, when the explicit analytical solution exists, as in the case of the TL model, the numerical integration can be performed using a finite energy range. However, when the exact $\varepsilon_1$ form is unknown, the truncated Kramers-Kronig integration[28] may become necessary.

One drawback of the proposed method is that the overall fitting quality is controlled by a single quantity of $\sigma_{crit}$. In particular, when $\sigma_{crit}$ increases in a low energy region, the SE fitting in the high energy range deteriorates, as mentioned above. In an optical interference regime observed typically in a low energy region of a thin-film structure, $\sigma$ tends to increase as an optical model adopted in a SE analysis is only an approximated structure and the imperfection of a film structure generally deteriorates fitting quality.[3,10,15] In such cases, the accurate determination of high-energy optical properties is hindered. The absolute value of $\sigma_{crit}$ can be varied along $E_{HA}$ to improve $\sigma$ in a high energy region, where the optical properties can be approximated by a simpler optical model of surface roughness layer/bulk layer due to a small light penetration depth.[15] Such a measure, however, complicates overall analysis.

In our analyses of the MoO$_x$ layers, a single roughness layer was assumed. In general, transparent oxides exhibit low $n$ and the influence of the surface roughness is not strong.[1] In contrast, in strong light absorbing materials, a single roughness layer may not be sufficient to describe the optical response, particularly when a geometrical roughness size ($S_{rough}$) exceeds the general restriction for the wavelength of the probe light (i.e., $S_{rough} > 0.1\lambda_{probe}$).[2,7] If absorbing materials with large roughness are analyzed, a multilayer roughness (or interface) model needs to be adopted.[2,3,7,15] Such advanced optical models can easily be incorporated in the above mentioned analyses.

Even though MoO$_x$ is a TCO, MoO$_x$ does not exhibit free carrier absorption. In more conventional TCOs, including In$_2$O$_3$:Sn (ITO) and ZnO, quite strong free carrier absorption is observed in a low energy region (~ 1 eV) when the carrier concentration is ~$10^{20}$ cm$^{-3}$, for example. To describe the free carrier absorption, the Drude model expressed by two parameters (i.e., amplitude and broadening factors) needs to be included.[2,16] In this case, the number of initial fitting parameters becomes nine (five for the TL peak, two for the structure, and two for the Drude model) and a further effort is necessary to shorten the time required for the grid search.

In the developed method, the multi-step zooming grid search has been adopted. However, the grid search is still a time-consuming and a rate determining step in our analysis and the computational time for the grid search should further be shortened. The two approaches that can be applied for a high-performance grid search include i) the adaptation of a multi-core cluster computer or ii) the application of advanced algorithms such as random search and Bayesian methods. Accordingly, more sophisticated schemes could be developed by advancing the automated SE analyses proposed here.

In this study, only a pair of ellipsometry spectra measured at one incident angle ($\theta = 55°$) was used for the analysis. More reliable SE analyses could be performed by incorporating multiple ellipsometry spectra measured at different incident angles and by further combining T/R spectra, as performed previously.[29]

When the depolarization of the incident probe light occurs by film thickness inhomogeneity and backside light reflection,[1] such non-ideal optical effects need to be accounted for properly. The influence of light depolarization can be incorporated into the developed automated method based on procedures established previously,[30-32] although the type of a depolarizing effect should be specified in advance in such cases.

In the developed method, SE analyses are performed by gradually increasing $E_{HA}$. However, the analysis could be implemented in a reversed energy direction (i.e., from $E_{max}$ toward lower energy). In this case, when the analyzed energy is sufficiently high and a thin-film sample can be treated as an opaque film, a simple optical model (i.e., a surface roughness/bulk model without a substrate) can be employed. Such an approach could be effective particularly for the determination of high-energy optical transition components. Nevertheless, these analyses are expected to show some difficulties in energy ranges where thin-film samples show optical interferences.

## VI. CONCLUSION

We have developed a fully automated SE analysis method that allows the complete determination of material dielectric functions in a full measured spectral range without any prior knowledge of material optical properties and initial guess parameters. The developed method consists of a multi-step grid search for finding initial guess parameters and the following SE fitting. The unique feature of the proposed method is that the initial SE fitting is restricted in a low energy region and the analyzed energy region is gradually expanded toward higher energy while introducing an additional transition peak whenever the fitting quality exceeds a critical value until the dielectric function is determined in a full spectral range. For the automated analysis, a reliable algorithm has been established and the dielectric functions of materials are analyzed specifically by employing Tauc-Lorentz transition peaks. The developed methods have been applied successfully to determine the optical properties of two different $MoO_x$ layers, one of which exhibits a strong mid-gap absorption. For the $MoO_x$ layer without the mid-gap states, the automated SE analysis results in a dielectric function expressed by four Tauc-Lorentz peaks and this analysis shows a better fitting quality, compared with a manual SE analysis performed assuming fewer transition peaks. The complete automation of a SE analysis is also possible for the $MoO_x$ layer with deep mid-gap absorption and the dielectric function which exhibits a dual absorption structure is determined. The full automation of SE analyses incorporating dielectric function modeling has never been performed previously and the developed method opens a novel way for SE as a more general characterization method that allows the accurate determination of material structures and optical properties in a short time.

**Supplementary Material**

The numerical TL parameters obtained for $MoO_x$-A using the automated (Table I) and the manual (Table II) methods shown in Figure 4 and also for $MoO_x$-B based on the automated (Table III) and the manual (Table IV) analyses of Figure 7 are summarized.

**Data Availability**

The data that support the findings of this study are available from the corresponding author upon reasonable request.

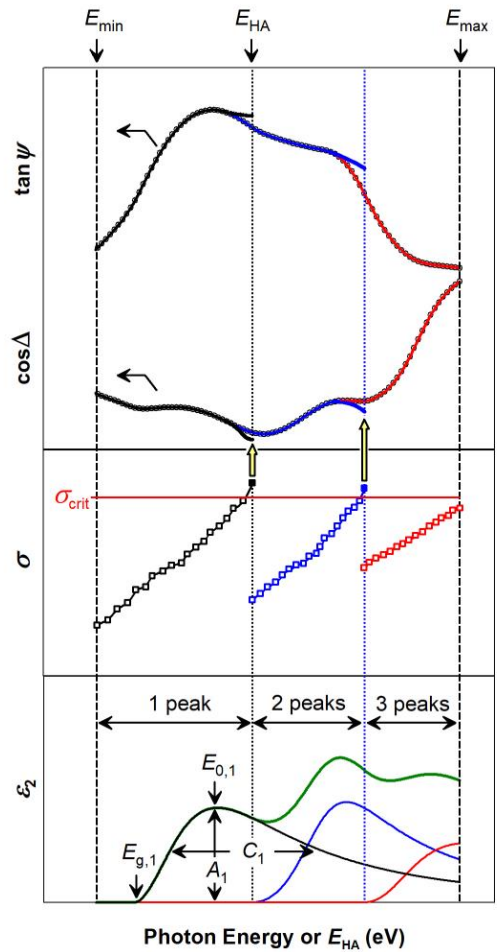

**FIG. 1.** Procedure of a fully automated SE analysis. In the proposed method, experimental $\tan \psi$ and $\cos \Delta$ ellipsometry spectra (open circles) are analyzed sequentially from a lower energy and the region of the SE fitting (solid lines) is expanded gradually toward higher energy. The SE fitting analysis is carried out using Tauc-Lorentz (TL) peaks assuming a simple optical model of roughness layer/bulk layer/substrate. In this example, the first stage of the SE analysis is performed at a lower energy using only one TL peak (black line). However, as the highest analyzed energy ($E_{HA}$) used in the SE analysis increases, the quality of the SE fitting deteriorates particularly in the high energy region as the sample optical properties cannot be expressed by the single TL peak. As a result, the fitting error $\sigma$ increases gradually with increasing $E_{HA}$. When $\sigma$ exceeds a critical value of $\sigma_{crit}$ (closed squares), an additional TL peak is incorporated to better express the complex optical properties of the sample and $E_{HA}$ is increased again (blue lines). This process is repeated until the dielectric function ($\varepsilon = \varepsilon_1 - i\varepsilon_2$) of the sample in a whole measured region ($E = E_{min} \sim E_{max}$) is determined. In the figure, the sample dielectric function is described by a total of three TL peaks and the resulting dielectric function is shown by the green line. The $E_{g,1}$, $E_{0,1}$, $A_1$, and $C_1$ indicate the $\varepsilon_2$ parameters of the first TL peak, which are band gap, transition energy, amplitude and broadening parameters, respectively.

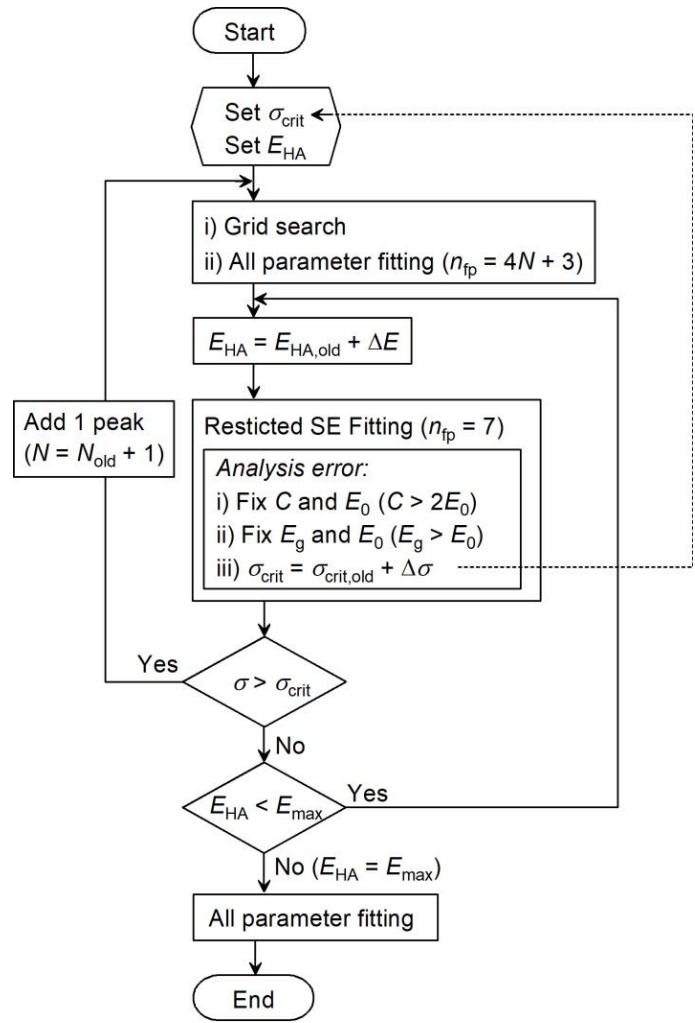

**FIG. 2.** Flow chart of the fully automated SE analysis developed in this study. This scheme has only two setting parameters of $\sigma_{crit}$ and $E_{HA}$. The $n_{fp}$ and $N$ show the numbers of free parameters used for the SE fitting and the TL peaks. The $E_{HA}$ is increased sequentially with an energy step of $\Delta E$. The restricted SE fitting indicates that only a TL peak with the highest transition energy is fitted with $n_{fp} = 7$ for $N > 1$. The analysis errors occur when the restrictions of the TL model ($C < 2E_0$ and $E_g < E_0$, for example) are violated. In such cases, the corresponding TL parameters are fixed or $\sigma_{crit}$ is increased by $\Delta\sigma$. When $E_{HA}$ reaches $E_{max}$, all parameter fitting is performed to obtain the final fitting result.

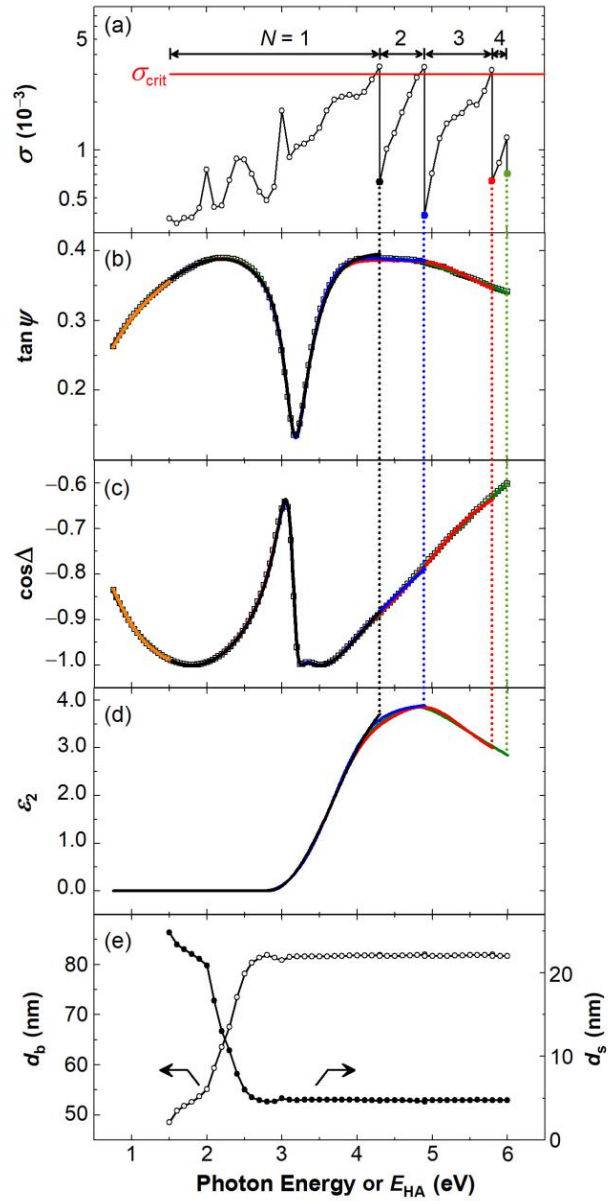

**FIG. 3.** SE analysis result obtained for the O-rich MoO$_x$ sample (MoO$_x$-A) using the fully automated method: (a) $\sigma$, (b) $\tan\psi$, (c) $\cos\Delta$, (d) $\varepsilon_2$, and (e) $d_b$ and $d_s$ values. The circles for $\sigma$ and ($d_b$, $d_s$) indicate the analysis results obtained for each $E_{HA}$ and the open squares for $\tan\psi$ and $\cos\Delta$ show the measured experimental spectra. For this sample, the optimum $\sigma_{crit}$ is $3 \times 10^{-3}$ and additional TL peaks are incorporated at $E = 4.3$ ($N = 2$), 4.9 ($N = 3$), and 5.8 eV ($N = 4$). The SE fitting results for $N = 2$, 3, and 4 at $E_{HA}$ indicated by the closed circles in (a) are shown by the black, blue and red lines. The green lines represent the final result obtained after implementing all parameter fitting at $E_{HA} = E_{max} = 6$ eV, whereas the orange lines indicate the SE fitting result derived from the zooming grid search and the following all parameter fitting with $E_{HA} = 1.5$ eV.

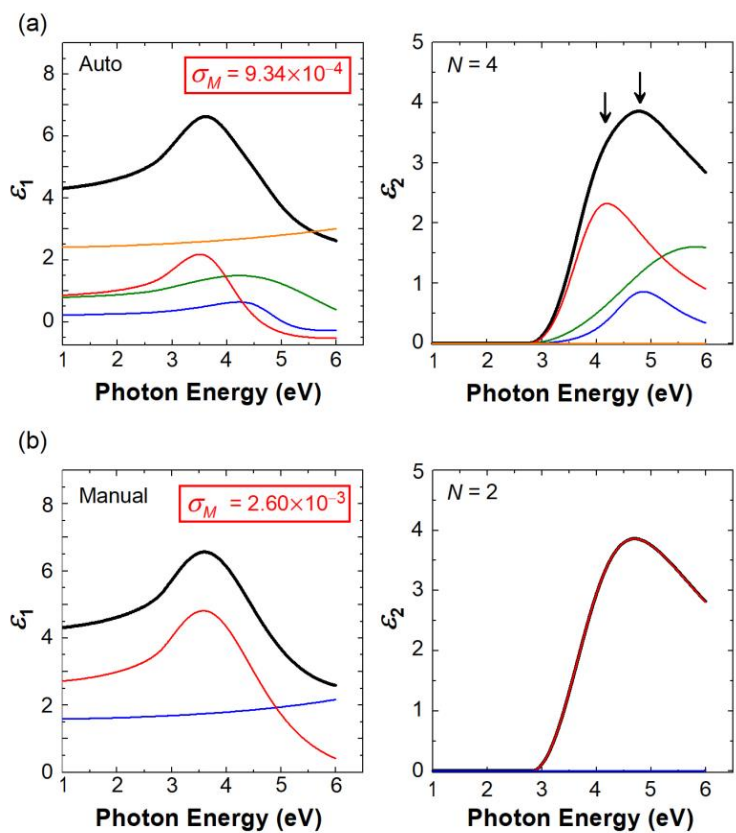

**FIG. 4.** Dielectric functions of MoO$_x$-A determined from (a) the automated SE analysis of Fig. 3 and (b) a manual SE analysis performed separately. The dielectric functions denoted by the black lines represent those of MoO$_x$-A deduced from the SE analyses and the contributions of each TL transition are indicated by colored spectra. The arrows in (a) indicate the shoulder peaks at $E = 3.87$ and $4.77$ eV. The $\sigma$ values for the whole spectra (i.e., $\sigma_M$) obtained in each analysis are also indicated. The automated analysis provides a better $\sigma_M$, compared with the manual SE analysis.

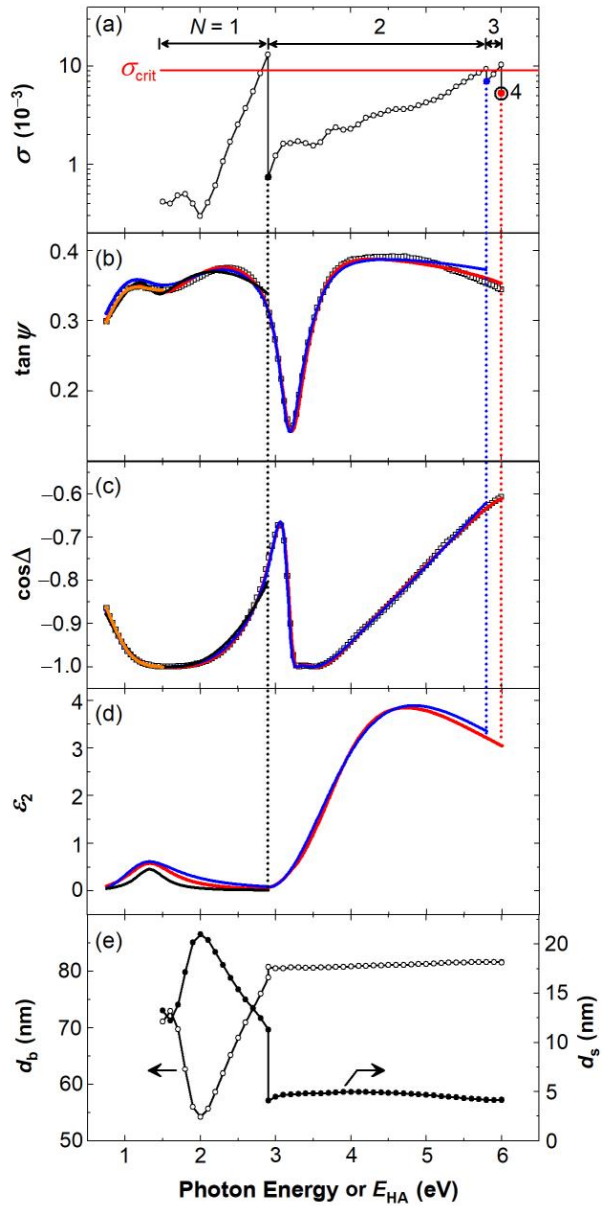

**FIG. 5.** Result of the automated SE analysis performed for the O-poor MoO$_x$ sample (MoO$_x$-B): (a) $\sigma$, (b) $\tan\psi$, (c) $\cos\Delta$, (d) $\varepsilon_2$, and (e) $d_b$ and $d_s$. The MoO$_x$-B exhibits a complex dielectric function due to the formation of the mid-gap states near 1.2 eV. The circles for $\sigma$ and ($d_b$, $d_s$) indicate the analysis results obtained for each $E_{HA}$ and the open squares for $\tan\psi$ and $\cos\Delta$ show the measured experimental spectra. The orange lines show the SE fitting result obtained from the zooming grid search and the following all parameter fitting with $E_{HA}$ = 1.5 eV. In this analysis, additional TL peaks are incorporated at 2.9, 5.8, and 6.0 eV and their SE fitting results indicated by the closed circles in (a) are shown by black, blue, and red lines, respectively.

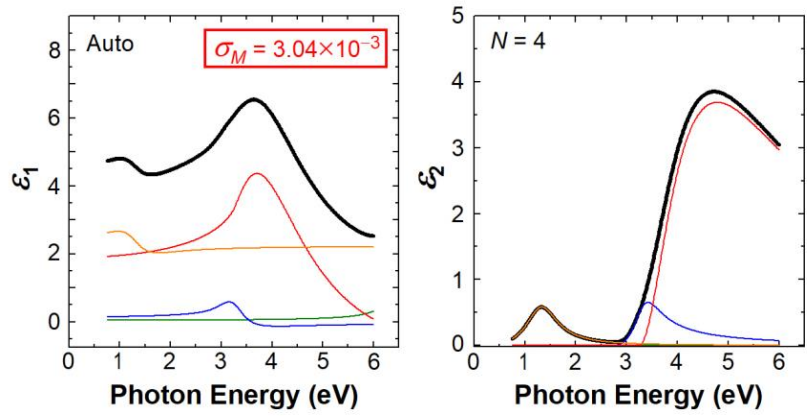

**FIG. 6.** Dielectric function of $MoO_x$-B obtained from the automated SE analysis of Fig. 5. The black lines indicate the dielectric function of $MoO_x$-B and the contributions of four TL transition peaks are indicated by colored spectra.

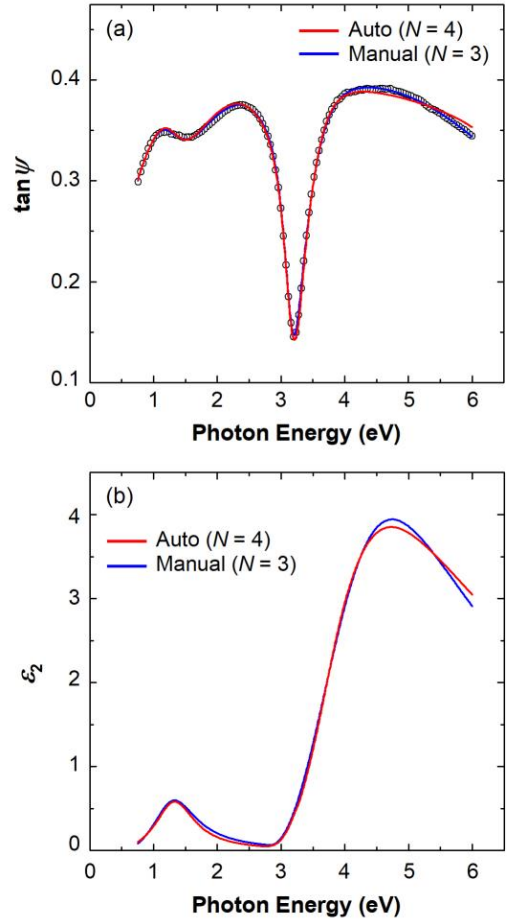

**FIG. 7.** (a) $\tan\psi$ spectra for $MoO_x$-B calculated by the automated SE analysis using four TL peaks (red line) and the manual SE analysis assuming three TL peaks (blue line), whereas the open circles show the experimental spectrum, and (b) $\varepsilon_2$ spectra of $MoO_x$-B obtained from the automated and manual analyses.

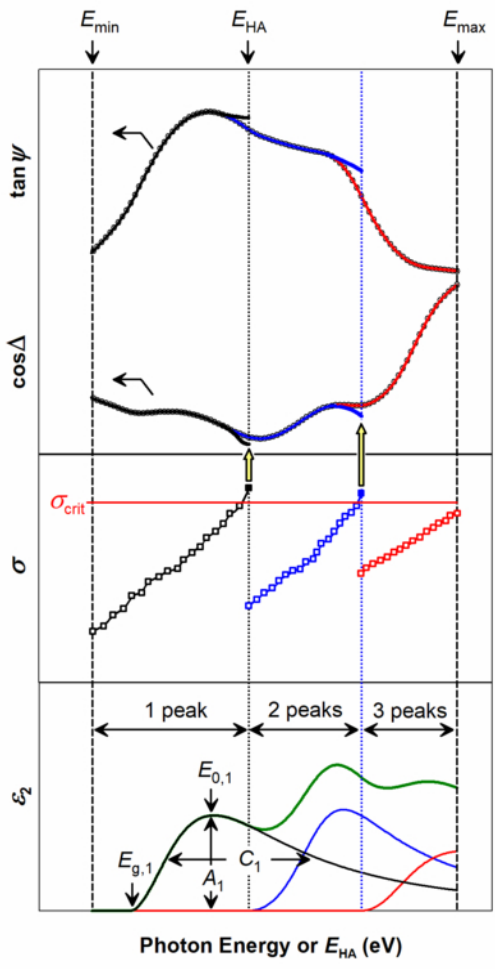

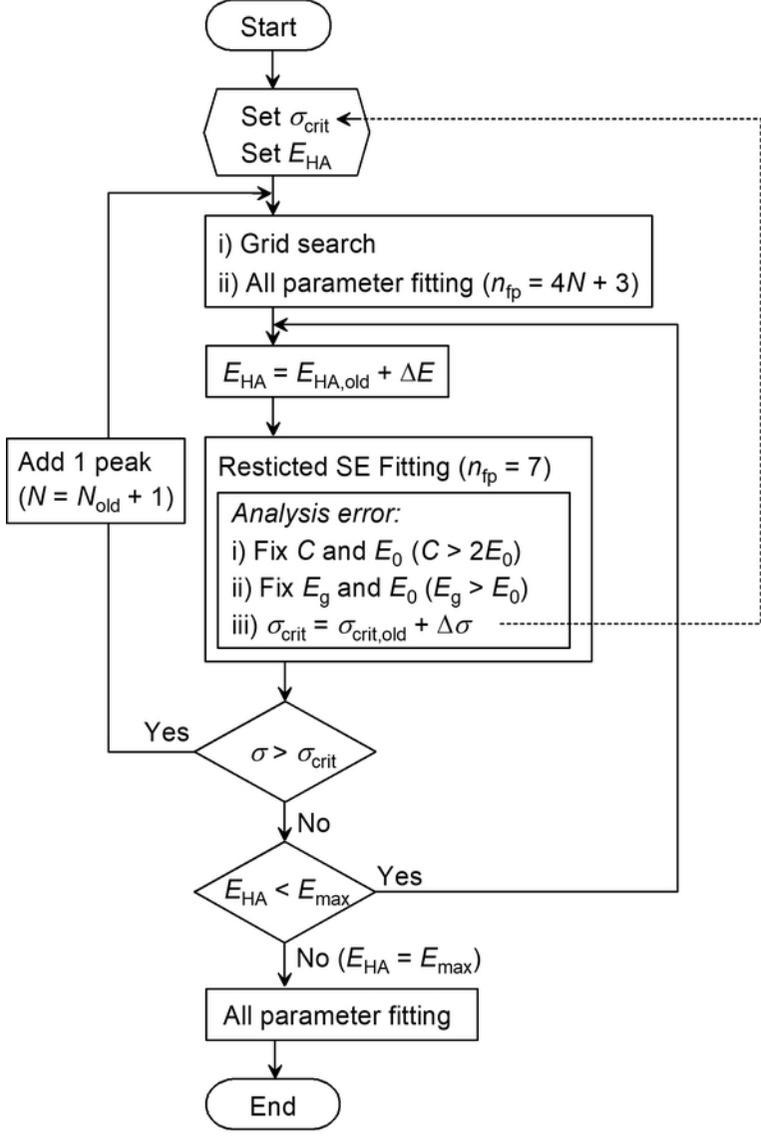

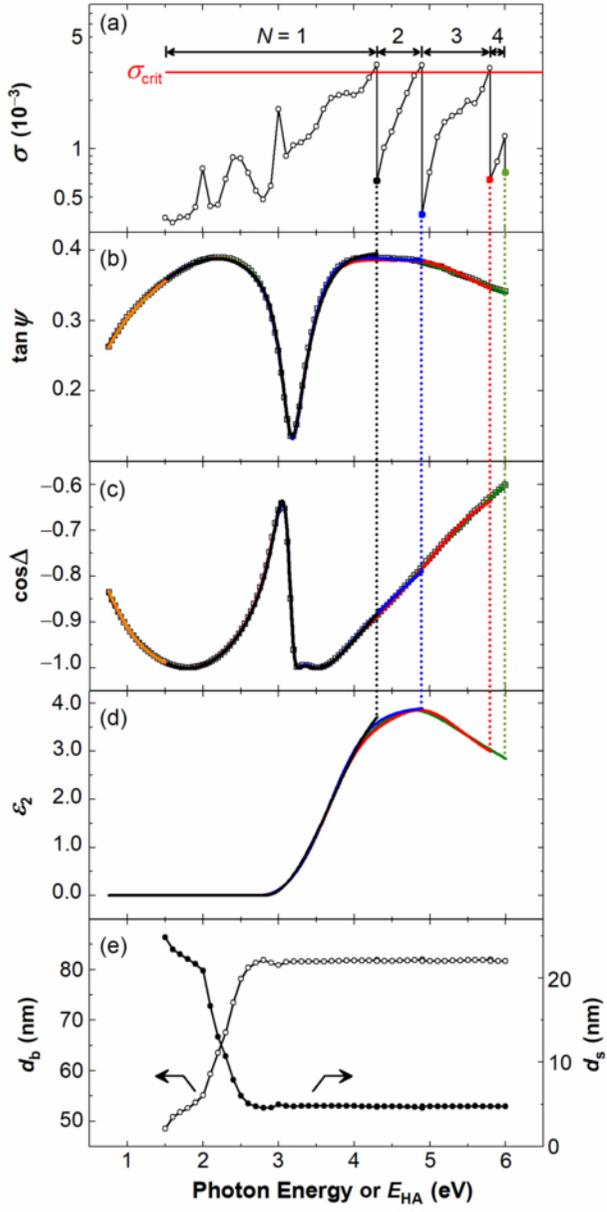

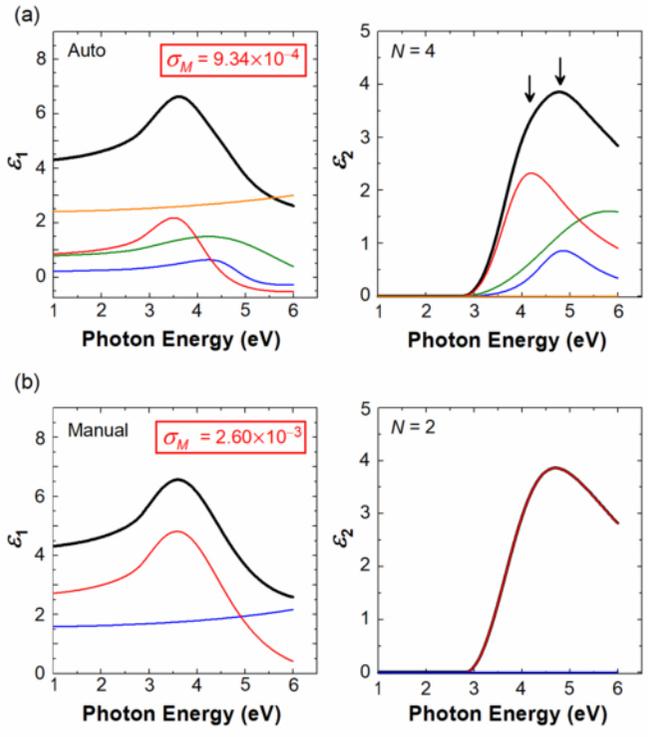

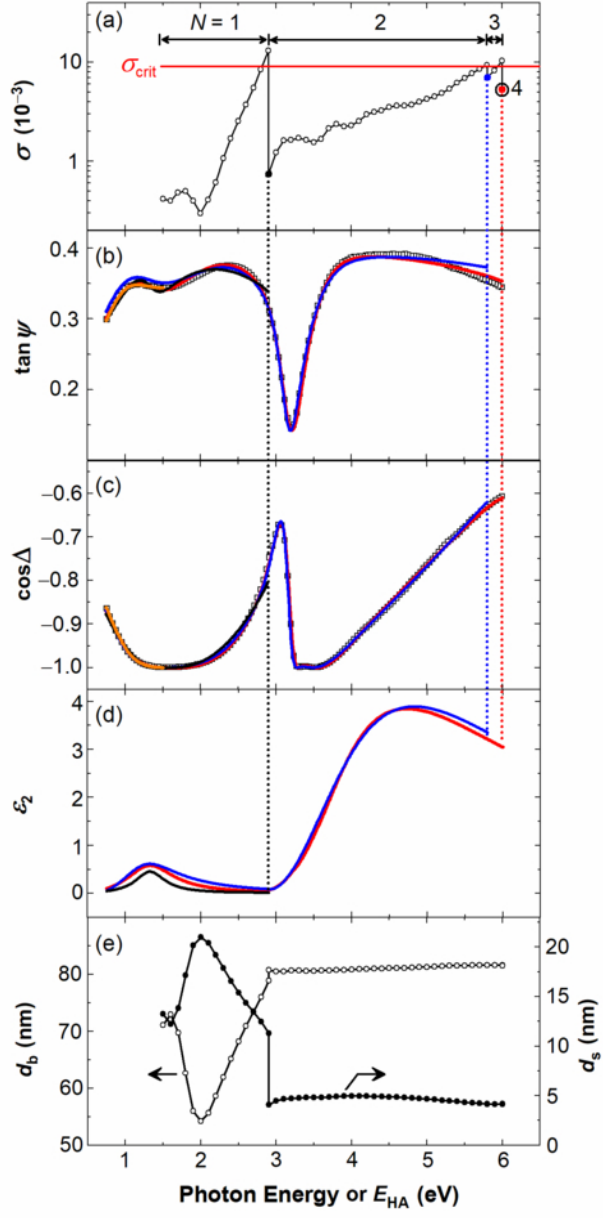

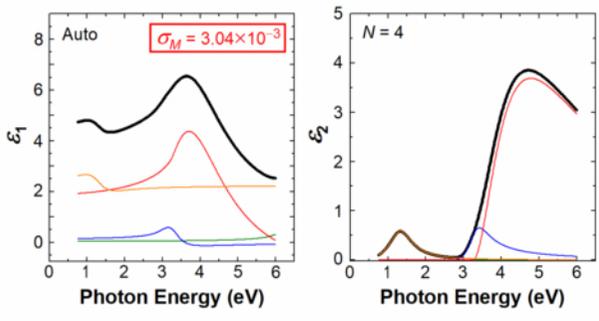

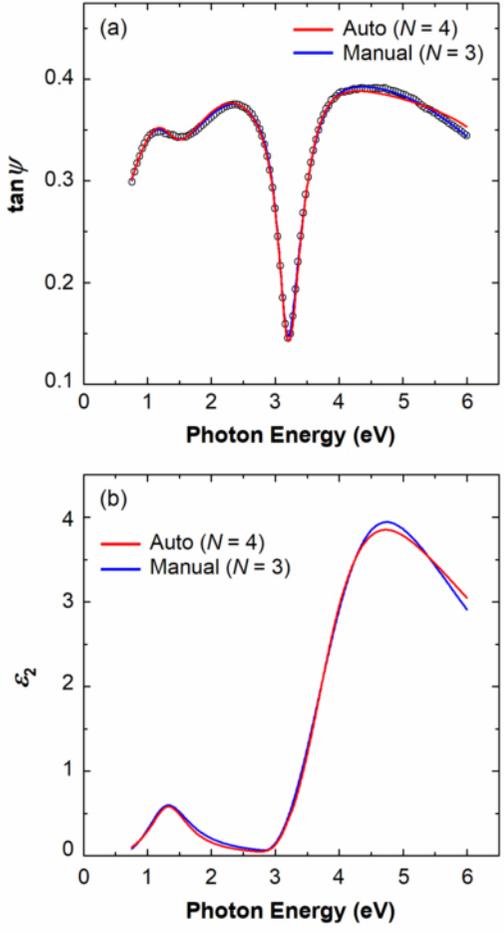